\pdfoutput=1
\documentclass[5p,final]{elsarticle}
\usepackage{fixltx2e}
\usepackage{amsmath,amssymb}
\usepackage{slashed}
\usepackage[latin1]{inputenc}

\def\alp{{(\alpha)}}
\def\bet{{(\beta)}}

\newcommand{\bfk}{\mathbf{k}}

\newcommand{\bfp}{\mathbf{p}}
\newcommand{\bfr}{\mathbf{r}}

\newcommand{\bfv}{\mathbf{v}}

\newcommand{\bfx}{\mathbf{x}}
\newcommand{\bfy}{\mathbf{y}}

\newcommand{\bfzero}{\mathbf{0}}

\newcommand{\barbfr}{\bar{\bfr}}

\newcommand{\dotp}{\boldsymbol{\cdot}}

\renewcommand{\div}{\operatorname{div}}

\newcommand{\ranstress}{s}
\newcommand{\RStress}{\tilde\tau^{(a)}}
\newcommand{\microstress}{\tau}
\newcommand{\CGStress}{\tau^{(a)}}
\newcommand{\lammpsstress}{\mathcal{T}}
\newcommand{\LAMMPS}{{\small LAMMPS}}
\newcommand{\pot}{\phi}
\newcommand{\bfg}{\mathbf{g}}

\journal{Chemical Physics}
\begin{document}
\begin{frontmatter}
\title{A numerical test of stress correlations in fluctuating hydrodynamics}
\author{Michael Schindler}
\address{Laboratoire~PCT, UMR~``Gulliver'' CNRS-ESPCI~7083, 10~rue Vauquelin, 75231~Paris~Cedex~05}
\ead{michael.schindler@espci.fr}
\begin{abstract}
The correlations of the fluctuating stress tensor are calculated in an equilibrium
molecular-dynamics simulation of a Lennard--Jones liquid. We define a
coarse-grained local stress tensor which can be calculated numerically and which
allows for the first time to determine the stress correlation function both in
time and in space. Our findings corroborate the assumptions made in fluctuating
hydrodynamics as long as the liquid is isotropic, that is in bulk. In the
vicinity of a rigid plate, however, the isotropy is restricted, and major
modifications must be done with respect to the usual theory. Among these are the
appearance of five different viscosities instead of two and a non-trivial
dependence of the distance from the wall. We determine these viscosities from
the simulation data and find that their values are very different from the bulk
values. We further find much longer relaxation times of the stress correlations
than in bulk.
\end{abstract}
\end{frontmatter}


\section{Introduction}% <<<

The theory of \emph{fluctuating hydrodynamics} starts with classical
hydrodynamics, which describes the average motion of a fluid and which is
formulated in terms of smooth tensor fields, and enriches it by additive
fluctuations. It can be understood as the attempt to bridge the gap between
microfluidics and nanofluidics from the large-scales side. The central question
is of course, what properties the fluctuations must have in order to correctly
represent the microscopic states of the liquid in a given problem? The standard
approach~\citep{LanLif59,FoxUhl70} is to split the full velocity and stress
tensor into a macroscopic (denoted by an overbar in the following) and into
fluctuating parts (denoted by a tilde). The linearised governing equations take
then the form of a two-scale process, with the standard compressible
Navier--Stokes equations for the macroscopic velocity~$\bar\bfv(\bfx,t)$ and the
corresponding Newtonian stress tensor~$\bar\sigma_{ij}(\bfx,t)$,
\begin{equation}
  \label{newtonstress}
  \bar\sigma_{ij} = -\delta_{ij}\bar p(\bfx,t) + \eta(\partial_i\bar
  v_j{+}\partial_j\bar v_i) +
  \Bigl(\lambda{-}\frac{2}{3}\eta\Bigr)\delta_{ij}\operatorname{div}\bar\bfv.
\end{equation}
Here and in the following, $\eta$~denotes the shear viscosity of the fluid and
$\lambda$~its volume viscosity. As we will be concerned only about equilibrium
situations, the macroscopic equations reduce to
\begin{equation}
  \label{NavStok_macro}
  \bar\rho(\bfx,t) = \rho_0, \quad
  \bar p(\bfx,t) = p_0, \quad
  \bar\bfv(\bfx,t) = \bfzero,
\end{equation}
with constants~$\rho_0$ and $p_0$. The fluctuating fields satisfy a linearised
Navier--Stokes equation with additional random noise~$\ranstress_{ij}$,
\begin{align}
  \label{NavStok_meso}
  \rho_0\partial_t \tilde\bfv &= \operatorname{div}(\tilde\sigma + \ranstress), \\
  \partial_t\tilde\rho &= -\rho_0 \div\tilde\bfv.
\end{align}
The stress tensor in equation~\eqref{NavStok_meso} is again split into two
parts, the first of which satisfies the same Newtonian
definition~\eqref{newtonstress} as the macroscopic stress tensor, only in terms
of a fluctuating velocity~$\tilde\bfv(\bfx,t)$ and of a fluctuating
pressure~$\tilde p(\bfx,t)$. The second contribution is an \emph{uncorrelated
fluctuating stress tensor}~$\ranstress_{ij}(\bfx,t)$. The tensor
$\tilde\sigma_{ij}$ is meant as a coarse-grained stress which is fluctuating,
but not as randomly as~$\ranstress_{ij}$, see for example the book by
\citet{KubTodHas85} for coarse-graining concepts, and Ref.~\citep{GraTalHan77}
for substitute Markov processes. The same two-step separation has been used by
\citet{HauMar73,ChoHer72}, but without commenting on why it is necessary.
Equation~\eqref{NavStok_meso} has the structure of a Langevin equation, with
$\ranstress_{ij}$ the random driving. This random stress tensor vanishes on
average, $\langle\ranstress_{ij}(\bfx,t)\rangle=0$, and it has to satisfy the
\emph{fluctuations--dissipation theorem of the second kind}. It is thus
uncorrelated in space and in time,
\begin{equation}
  \label{bulkcorr}
  \bigl\langle\ranstress_{ij}(\bfx,t)\,\ranstress_{kl}(\bfy,s)\bigr\rangle
  = 2kT\,A_{ijkl}\,\delta(\bfx{-}\bfy)\,\delta(t{-}s).
\end{equation}
Here, $\langle\cdot\rangle$~denotes the average over different realisations, or
a time average; $k$~is Boltzmann's constant, and \textit{T}~the temperature of
the system. The correlator is further assumed to be isotropic, which fixes the
fourth-rank tensor~$A_{ijkl}$ to~\cite{FoxUhl70}
\begin{equation}
  \label{isotropy}
  A_{ijkl} = \eta\bigl[\delta_{ik}\delta_{jl} + \delta_{il}\delta_{jk}\bigr]
  + \Bigl(\lambda - \frac{2}{3}\eta\Bigr) \delta_{ij}\delta_{kl}.
\end{equation}
This special form of~$A_{ijkl}$ follows from the fact that it must be isotropic, thus
invariant under any rotation of the coordinate system. It is therefore a linear
combination of the only three isotropic tensors of rank four, which are
$\delta_{ij}\delta_{kl}$, $\delta_{ik}\delta_{jl}$, and
$\delta_{il}\delta_{jk}$~\cite{Temple60}. The factors of proportionality are
found to be the three viscosities of a fluid~\cite{GroMaz84}. The additional
symmetry of the stress tensor, leading to $A_{ijkl} = A_{jikl} = A_{ijlk}$
eliminates the rotational viscosity.

In the vicinity of a rigid boundary, however, isotropy is not granted in the
same way as in the bulk. Instead, the stress correlator should reflect the fact
that the normal direction is different from the two tangential directions. In
this paper, we therefore ask which tensorial form the correlator will adopt
when evaluated close to a rigid wall. In order to start with the most simple
situation, we regard the stress correlator in thermodynamic equilibrium.

We must further ask, what ``close'' means, that is up to which distance from the
wall the different symmetry is sensed by the fluctuating part of the stress
tensor. Near the boundary, we thus question also the spatial delta function in
Eq.~\eqref{bulkcorr}. The problem with the spatial delta function becomes
apparent when applying fluctuating hydrodynamics for example to the Brownian
motion of a spherical particle. This has been done in the literature on the
autocorrelation of a Brownian particle~\cite{HauMar73,ChoHer72}. If we naively
integrate the random stress tensor over the particle in order to obtain the
random force, and then construct the autocorrelation of that force, the result
would always be infinite. The reason is that we then integrate the
three-dimensional delta function from Eq.~\eqref{bulkcorr} only over the
two-dimensional surface of the particle. The third dimension thus remains a
delta function, evaluated at zero.

The spatial delta function must be questioned also from a less formal point of
view: The standard argument for the absence of spatial correlations is based on
thermodynamics~\citep{LanLif59}: The fluid volume is cut into several small but
finite portions~$\Delta V$. In each of these volumes, the entropy production is
calculated as the volume-integral of the tensorial reduction of the stress with
the rate-of-strain. The fluctuations of the stress tensor is then taken from the
variations of the entropy production around the equilibrium state. As this
integral does not contain any surface contribution, the fluctuations of two
stress tensors attributed to two such boxes are independent of each other. This
property is then kept \textit{even in the limit of infinitesimal box size}. It
appears as if there should be a lower spatial bound to the validity of this
argument, namely when the volume size become of the order of the distance
between the molecules of the liquid. As thermodynamic arguments are used, the
thermodynamic limit of many interacting molecules is assumed to be done before
the box size can be sent to zero. It appears thus as a reasonable question to
ask, down to which spatial scale does the delta function yield a good
description? It will turn out in the following, that it works amazingly well: In
bulk fluid, the delta function holds down to distances smaller than the
intermolecular spacing.

It is the aim of this paper to test the validity of Eqs.~\eqref{bulkcorr}
and~\eqref{isotropy}, describing stress correlations---in particular, to find
deviations in the vicinity of a rigid surface. In order to do so, we start from
a fully microscopic simulation of a liquid and extract the stress tensor using
coarse-graining. As the simulation is done at thermal equilibrium, the
macroscopic fluid velocity~$\bar\bfv$ vanishes, leaving only the thermodynamic
pressure in the macroscopic stress tensor~$\bar\sigma_{ij}=-p_0\delta_{ij}$. We
correlate the coarse-grained stress tensor at different positions and at
different times in order to verify both the isotropy in Eq.~\eqref{isotropy} and
the delta-functions in Eq.~\eqref{bulkcorr}. This will be done both near a rigid
wall and in bulk fluid.

The precise form of the stress tensor correlations is potentially important for
all fluidic systems including boundaries. In microfluidic applications, and even
more in nano\-fluidic ones, the role of the boundaries is
essential~\citep{KetReiHanMul00,GutEtal04,SchTalKosHan07,JohHanThi08}. In
nanofluidics, the presence of boundaries even introduces different time scales
in the relaxation dynamics~\citep{BocCha10}. For example, Brownian motion of a
particle immersed in a viscous fluid, which is one of the most classical systems
both in the hydrodynamic and in the stochastic community, is in most cases a
microfluidic problem---depending on the size of the particle. Both for large and
for small particles, the velocity fluctuations are known to exhibit long-time
tails in their autocorrelation-function. The theoretical explanations of this
phenomenon, as far as they consider both the particle and the fluid, assume the
isotropic form of Eqs.~\eqref{bulkcorr} and~\eqref{isotropy} for the fluctuating
part of the stress tensor---even when this stress tensor is integrated over the
surface of the particle to yield the total fluctuating force acting on
it~\cite{HauMar73,ChoHer72}. The investigation in the present paper is one of
two necessary steps to render the theoretical description of the long-time tails
more consistent. We will find below that it is not the usual viscosity which
occurs in the formula of the force autocorrelation, but a viscosity which is a
property of the fluid and the structuring effect of the rigid particle itself.
The other step will be published in Ref.~\citep{Schindler10}.

We will further find that the inclusion of boundaries makes it necessary to pass
from homogeneous viscosities (which are defined in the limit of homogeneous
space without any scale) to rather \emph{local} viscosities. Such viscosities
reflect more details of the geometrical conditions and of the time scales
involved.
% >>>

\section{Symmetry considerations}% <<<
\label{sec:symmetry}

The form~\eqref{isotropy} of the stress correlator follows directly from
symmetry considerations. We will now apply the same considerations to derive the
corresponding expression for a liquid close to a wall. Similar formulae are
found in the literature on nematic liquids~\citep{BohBraPle04}. The following
symmetries apply to the stress correlator:

(A) The most rigorously enforced symmetry is that the \emph{stress tensor is
symmetric}. This symmetry is inherited from the microscopic definition of the
stress tensor and is thus not subject to random fluctuations. It leads to the
conditions
\begin{equation}
  \label{symmetry}
    \langle\ranstress_{ij} \ranstress_{kl}\rangle
  = \langle\ranstress_{ji} \ranstress_{kl}\rangle
  = \langle\ranstress_{ij} \ranstress_{lk}\rangle.
\end{equation}
By this symmetry, the~$81$~components in the full correlation tensor of rank
four in three-dimensional space are reduced to $36$~different ones.

(B) Another symmetry is \emph{isotropy}, that is the invariance of the
correlator under rotations. It is not enforced by the microscopic definition and
must result from the averaging either over an ensemble or possibly over a long
trajectory. In bulk, the rotations are arbitrary, while in a geometry with
plates one direction is distinct from the two others and only rotations parallel
to the plates are permitted. The maximal number of different components is now
reduced to~$21$. They are, however, not all completely independent, and some of
them vanish: A systematic decomposition of the rank-four tensor into parallel
and orthogonal parts, using parallel isotropy and Eqs.~\eqref{symmetry}, reveals
that these $21$~entries can take only nine~different non-zero values, and that
they depend on seven~independent parameters (viscosities).

(C) The next symmetry is the invariance under \emph{index pair exchange},
\begin{equation}
  \label{pairexchange}
    \langle\ranstress_{ij} \ranstress_{kl}\rangle
  = \langle\ranstress_{kl} \ranstress_{ij}\rangle.
\end{equation}
It is a consequence of the entropy production being a scalar~\citep{GroMaz84}.
This symmetry leads to the identification of two of the viscosities and the
vanishing of another. After having used all symmetry requirements, we thus have
a total of five~viscosities left, which give rise to six~different non-zero
values in the $13$~principally different components of the stress correlator.%
\begin{figure}%
  \centering
  \includegraphics{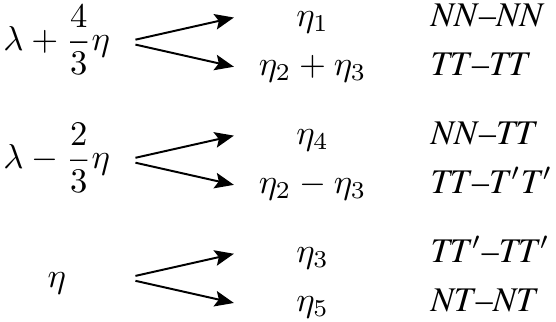}%
  \caption{Schematic breaking of viscosities in bulk fluid (left) into the
  surface viscosities near a flat wall (right). Given are the three/six non-zero
  values found in the components of the stress tensor correlator~$A_{ijkl}$,
  indexed by the normal direction~\textit{N} and the two tangential directions
  \textit{T} and~\textit{T}$^\prime$.}%
  \label{fig:viscosities}%
\end{figure}%

The symmetry conditions near a plate require remarkably more viscosities than
those in bulk, where two~viscosities lead to three~different values. The
presence of the wall thus induces a separation of the bulk viscosities into
surface viscosities. This splitting is depicted in Fig.~\ref{fig:viscosities}.
The diagram provides on the left-hand side the three (in bulk) different values
found in the components of Eq.~\eqref{isotropy}, and on the right-hand side the
six non-zero values near a flat rigid immobile plane. The full rank-four tensor
expressing the random stress correlations near a wall is
\begin{multline}
  \label{surfiso}
  A_{ijkl} = \eta_1\, N_i N_j N_k N_l
  + \eta_2\: g_{ij}g_{kl} \\
  + \eta_3\, (g_{ik}g_{jl}{+}g_{il}g_{jk}{-}g_{ij}g_{kl})
  + \eta_4\, (N_iN_jg_{kl}{+}N_kN_lg_{ij}) \\
  + \eta_5\, (N_iN_kg_{jl}{+}N_jN_kg_{il}{+}N_iN_lg_{k}{+}N_jN_lg_{ik}),
\end{multline}
with the shortcut~$g_{ij} := \delta_{ij}-N_iN_j$.

The rightmost column of Fig.~\ref{fig:viscosities} uses a symbolic notation
for the indices of tensor components. \textit{N}~stands for the direction normal
to the wall, \textit{T}~may stand for both tangent directions, and \textit{T\/$'$}
is only used if both tangent directions are used: In particular,
\textit{TT}~stands for the two index combinations with the same tangent vector,
whereas $\textit{TT\/}'$~stands for the ones with different tangent vectors. The
same notation will be used in the figures of Sec.~\ref{sec:md} to visualize the
$13$~components of the stress correlation tensor, six~of which are non-zero and
seven~vanish.
% >>>

\section{Coarse-graining of a microscopic stress tensor}% <<<

Let us consider a molecular dynamics simulation in order to study the
microscopic dynamics of the molecules in a liquid. The liquid here consists of
point-like atoms, the dynamics of which is governed by Hamilton's equations
comprising a pair-interaction potential. In this dynamics we define a
\emph{microscopic stress tensor}. It is chosen such that its
\emph{coarse-grained version}, which is obtained by integrating it over a small
region, will be consistent with a macroscopic notion of the hydrodynamic stress
tensor.

The Hamiltonian of \textit{N}~point-like atoms~$\alp$ of masses $m^\alp$,
positions~$\bfr^\alp(t)$ and linear momenta~$\bfp^\alp(t)$, which undergo mutual
influence via a pair-potential~$\pot$, reads
\begin{equation}
  H = \sum_{\alpha=1}^N \frac{\bfp^\alp\dotp\bfp^\alp}{2m^\alp}
  + \sum_{\alpha=1}^N \sum_{\beta=1}^{\alpha{-}1}
    \pot\bigl(\|\bfr^\bet{-}\bfr^\alp\|\bigr).
\end{equation}
The time evolutions of the positions and momenta are given by Hamilton's
equations. The microscopic stress tensor~$\microstress_{ik}$, can then be
introduced as the current density of the \emph{density of linear
momentum}~$\bfg$, reading
\begin{equation}
  g_i(\bfr, t) := \sum\nolimits_{\alpha} p^\alp_i(t)\: \delta\bigl(\bfr - \bfr^\alp(t)\bigr).
\end{equation}
In the absence of external forces, the conservation of momentum is expressed by
the following balance equation, which introduces the microscopic stress
tensor~$\microstress_{ik}$,
\begin{equation}
  \label{momcons}
  \partial_t g_i(\bfr, t) - \partial_{r_k} \microstress_{ik}(\bfr, t) = 0.
\end{equation}
The definition of these tensor fields allows us to pass from a description in
terms of the atom positions~$\bfr^\alp(t)$ to a field description for any
position~$\bfr$. Equation~\eqref{momcons}, which is at the same time a
conservation theorem and a definition, fixes~$\microstress$ only partially. We
may always add an arbitrary part with zero divergence, a gauge freedom which is
inherent to the continuum description. However, we will not go into the
implications of this freedom. A convenient definition of the microscopic stress
tensor is the following, since it is symmetric and lends itself to a
straight-forward coarse-graining technique,
\begin{multline}
  \label{forster:stress}
  \microstress_{ik}(\bfr, t) = -\sum_\alpha
    \frac{p_i^\alp\, p_k^\alp}{m^\alp}
    \delta\bigl(\bfr{-}\bfr^\alp(t)\bigr) \\
   {}+ \frac{1}{2}\sum_\alpha \sum_{\beta\neq\alpha}
    \pot^\prime\bigl(\|\bfr^\bet{-}\bfr^\alp\|\bigr)
    \frac{r_i^\alp{-}r_i^\bet}{\|\bfr^\bet{-}\bfr^\alp\|} \times\\
    \bigl(r_k^\alp{-}r_k^\bet\bigr)
    \int\limits_0^1 \delta\Bigl(\bfr{-}\bfr^\alp{-}\lambda\bigl(\bfr^\bet{-}\bfr^\alp\bigr)\Bigr)
    \:d\lambda.
\end{multline}
The first term is usually called the \emph{kinetic} part of the stress tensor.
The second term comprises virial contributions, together with the integral of a
three-dimensional delta function over the line starting at the point~$\bfr^\alp$
and ending at~$\bfr^\bet$. Expression~\eqref{forster:stress} is the same as that
of~\citet[see Eq.~(4.6)]{Forster90}, with the exception of the lower boundary of
this integral.

We choose to do a coarse-graining by averaging the microscopic stress tensor
within a small cube~$V^a(\bfx)$ of side-length~$a$, centred around~$\bfx$. We
call this average the \emph{coarse-grained stress
tensor}~$\CGStress_{ik}(\bfx,t)$ at the point~$\bfx$,
\begin{equation}
  \CGStress_{ik}(\bfx,t) := \frac{1}{a^3} \int\limits_{V^a(\bfx)}
  \microstress_{ik}(\bfr,t)\:d\bfr.
\end{equation}
The integration of the kinetic term in $\microstress_{ik}$ is readily done. It
uses the momenta of the particles which are found within the volume at time~$t$.
The line integral of the delta function has a simple geometrical interpretation.
When integrated over the volume~$V^a(\bfx)$, only those parts of the line
integral contribute where the line is found within the volume. For the (convex)
cubic box~$V^a(\bfx)$, only four different cases are possible:
\begin{multline}
  \label{intcases}
    \|\bfr^\bet{-}\bfr^\alp\|
    \int\limits_V
    \int\limits_0^1 \delta\Bigl(\bfr{-}\bfr^\alp{-}\lambda(\bfr^\bet{-}\bfr^\alp)\Bigr)
    \:d\lambda\:d\bfr \\{} = \left\{\begin{array}{cl}
      \|\bfr^\bet{-}\bfr^\alp\| & \text{if $\bfr^\alp\in V$ and $\bfr^\bet\in V$,} \\
      \|\barbfr{-}\bfr^\alp\| & \text{if $\bfr^\alp\in V$ and $\bfr^\bet\notin V$,} \\
      \|\bfr^\bet{-}\barbfr\| & \text{if $\bfr^\alp\notin V$ and $\bfr^\bet\in V$,} \\
      \|\barbfr{-}\barbfr'\| & \text{if $\bfr^\alp\notin V$ and $\bfr^\bet\notin V$.}
    \end{array}\right.
\end{multline}%
\begin{figure}
  \centering
  \includegraphics{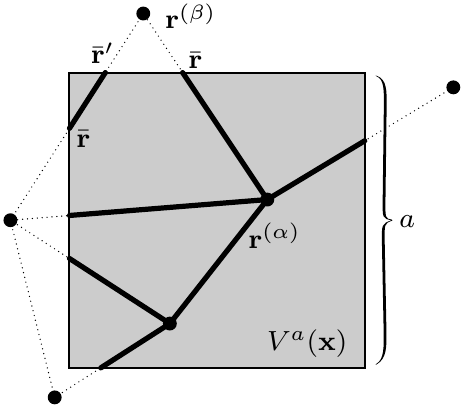}%
  \caption{Visualisation of the virial contributions to the coarse-graining of
  the stress tensor. From the connection lines between the atoms (small circles)
  only the bold parts are taken into account. All cases of Eq.~\eqref{intcases}
  are displayed.}%
  \label{fig:cases}%
\end{figure}
Here, $\barbfr$ and~$\barbfr'$ denote the points where the connection line
between the atoms intersect the boundary of the volume~$V^a(\bfx)$. The last
case in Eq.~\eqref{intcases} yields no contribution if these intersection points
do not exist. Figure~\ref{fig:cases} provides a graphical representation of
these different cases. For calculating the coarse-grained stress tensor we thus
have to find the contributing atom pairs, to evaluate the pair potential, and to
find the intersection points. Notice that even a box which does not contain any
particle may yield a non-zero microscopic stress. It is sufficient that any
connecting line passes through the box.

The usual way to calculate a stress tensor in molecular dynamics simulations is
to add up the kinetic contribution and the full virial of the atoms. The
simulation package~\LAMMPS~\cite{lammps}, for example, outputs the following
quantity as \texttt{stress per atom},\footnote{Throughout the paper, terms
specific to~\LAMMPS\ are typeset in a typewriter font.}
\begin{multline}
  \label{lammps:stress}
  \lammpsstress^{(\alpha)}_{ik} := - m^{(\alpha)} v^{(\alpha)}_i v^{(\alpha)}_k
  + \frac{1}{2} \sum_{\beta\in\text{Nbrs.}(\alpha)}
    (r_k^\alp{-}r_k^\bet)\times{}\\
    \frac{r_i^\alp{-}r_i^\bet}{\|\bfr^\alp{-}\bfr^\bet\|}\:
    \pot^\prime\bigl(\|\bfr^\alp{-}\bfr^\bet\|\bigr).
\end{multline}
However, $\lammpsstress^{(\alpha)}_{ik}$~has the dimension of stress times
volume. The stress is attributed to the atom with index~$\alpha$, its neighbours
have indices~$\beta$. It is clear that $\lammpsstress^{(\alpha)}_{ik}$~cannot be
a local quantity, as it does not depend on any space variable. It may be the
integral of a local quantity---such as the stress in
Eq.~\eqref{forster:stress}---over a specific volume around the
atom~$\alpha$---but over which volume? We find the answer to this question from
the explicit averaging procedure above: Only if the particle and all its
neighbours are within the volume of integration, that is the first case in
Eq.~\eqref{intcases}, then we obtain expression~\eqref{lammps:stress} as the
contribution coming from this atom. We conclude that
$\lammpsstress^{(\alpha)}_{ik}$~is the bulk contribution to a stress tensor when
integrated over a volume which is sufficiently large that it contains the atom
in question and all its interaction partners. It discards all boundary terms.
For geometrical reasons it is not possible to assign space-filling individual
volumes to all atoms (such as is done for example by a Voronoi tessellation) in
order to obtain the stress per atom as an integral over the local stress. As the
cell boundaries in the Voronoi tessellation are determined only by nearest
neighbour's positions, they cannot correctly represent all interaction partner's
positions, which are spanned over a larger region.

As our interest lies in the \emph{local} fluctuations of the stress tensor, we
have to keep the integration boxes as small as possible. The small boxes forbid
the use of Eq.~\eqref{lammps:stress}, we have to average the microscopic stress
tensor explicitly, according to the cases in Eq.~\eqref{intcases}. The
additional work to be done in comparison with Eq.~\eqref{lammps:stress} is the
solution of the geometrical intersection problem.
% >>>

\section{Results from molecular dynamics simulations}% <<<
\label{sec:md}

The molecular dynamics simulations described in the following have been done
with the molecular dynamics package~\LAMMPS~\citep{lammps}. The coarse-graining
procedure described above has been added as a module.
% >>>
\subsection{Details of the simulation}% <<<

\begin{figure}
  \centering%
  \includegraphics{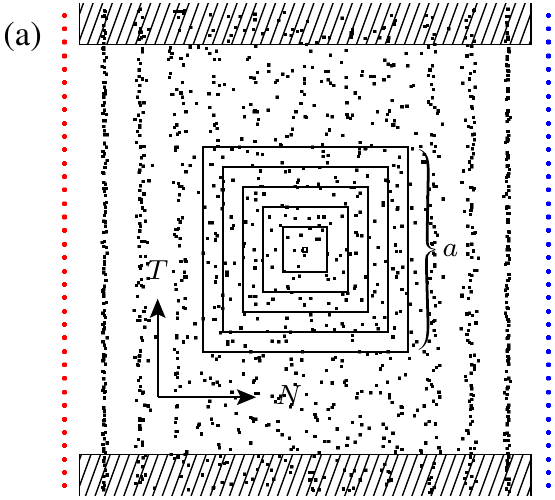}\\
  \includegraphics{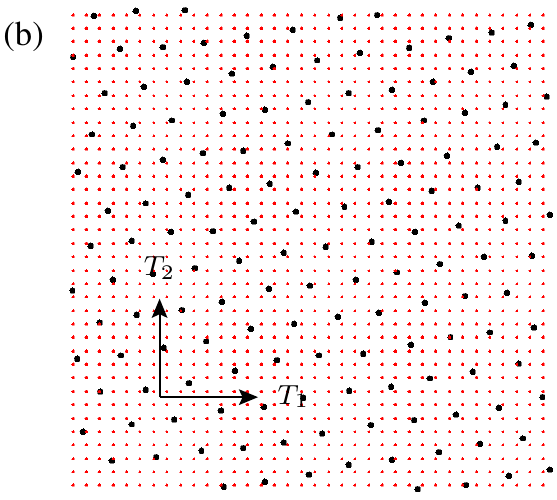}%
  \caption{(Colour on-line) Snapshots of the simulation domain with $N=1380$
  fluid atoms~(black), confined between two fixed plates which consist of a
  regular grid of immobile atoms~(red/blue). Shown are cuts through the domain
  as seen from two different sides. The series of cubes in the middle of
  panel~(a) indicate the cubic boxes used for averaging in Sec.~\ref{sec:size}.
  In the hatched region, a Langevin thermostat controls the temperature.}%
  \label{fig:atoms}%
\end{figure}

Our simulation is set up as a three-dimensional Lennard--Jones liquid between
two plates, as depicted in Fig.~\ref{fig:atoms}. 1380~fluid atoms are confined
between two fixed plates. The plates are constructed as a single regular layer
of atoms with the same interaction potential. In order to approximate a smooth
flat surface, the lateral distance between these plate atoms is about a third of
that in the fluid. The simulation domain is periodically continued in the two
directions parallel to the plates.

The fluid atoms are advanced in time using the Verlet algorithm
(\texttt{NVE~integration}) which is consistent with the microcanonical ensemble.
An exception are the hatched regions in Fig.~\ref{fig:atoms}a, where an
additional Langevin thermostat is applied. Such a thermostat is necessary to
counteract numerical integration errors, but it modifies the statistics of the
stress. We therefore chose to apply it only in regions where we do not measure
the stress. The plate atoms are not advanced in time, simply by not applying any
integration routine. The forces on these atoms are calculated, however, which is
equivalent to the possible inclusion of strong constraining forces from
crystalline layers behind. This procedure allows to determine the force exerted
on the plates by the fluid.

The pair-interaction potential between the atoms is the $12/6$~Lennard--Jones
potential, modified by an interpolation term to have a finite carrier,
\begin{equation}
  \label{LJsmooth}
  \phi(r) = \left\{\begin{array}{cl}
    4\epsilon\bigl[(\sigma/r)^{12} - (\sigma/r)^6\bigr] & \text{for $r<2.5\sigma$}\\
    0 & \text{for $r>3.5\sigma$}\\
    \sum\limits_{n=-3}^{2} \xi_n (\sigma/r)^{2n} & \text{else.}
  \end{array}\right.
\end{equation}
The six parameters $\xi_n$~are determined to guarantee continuous derivatives
everywhere, up to the second. We use this smooth potential in order to make sure
that discontinuities in the second derivative do not produce artefacts in the
measurement of the stress tensor.

In the following, we will use \texttt{Lennard-Jones units} for all quantities.
In these units, distances are measured as multiples of the length
parameter~$\sigma$ in the Lennard--Jones potential, energy as multiples
of~$\epsilon$, and time in terms of $\sqrt{m\sigma^2/\epsilon}$, where $m$~is
the mass unit. In these units, the mass of the atoms is chosen to be~$1$, the
side-length of the simulation domain is fixed to be~$12$. The time unit is
approximately the oscillation period in the minimum of the pair
potential~$\phi$. The time-step of integration, used in the Verlet algorithm
is~$0.002$, thus more than two orders of magnitude smaller than all physical
time scales in the system.

The density and the (kinetic) temperature of the system are chosen such that it
is found in its liquid state, according to the phase diagram in
Refs.~\citep{Smit92,Hoef00,MasPab07}. Of course, this holds only in the ``bulk''
region where the structuring effect of the walls is absent. The disordered fluid
is visible in the middle of Fig.~\ref{fig:atoms}a. There, the following average
values for temperature, density and pressure are found: $T=1.04$, $\rho_0=0.83$,
$p_0=1.3$. Near the plates, the enforced flat geometry imposes a partial ordering
of the fluid. Up to three distinct layers of fluid atoms can be identified near
each wall. The first layer has a locally hexagonal structure as seen in
Fig.~\ref{fig:atoms}b. Its width and orientation are independent of the grid
spacing of the plate atoms. Due to periodic boundary conditions, the hexagonal
grid does not rotate anymore as soon as the system is equilibrated. Of course,
also the stress tensor becomes anisotropic close to the plates.

The simulation was first run to let the system find its thermodynamic
equilibrium. We then continued the simulation while calculating the
coarse-grained stress tensors at some fixed positions every $5$~time steps.
These time series are analysed in the following, either using the blocking
method~\citep{FlyPet89} to ensure that all used variables are equilibrated, or
by simple averaging to measure spatial correlations of the stress tensor, or by
fast Fourier transforms for the time correlations. The
averages~$\langle\cdot\rangle$ in the following are thus time-averages over a
long trajectory. As the system is in equilibrium, this should be equivalent to
an ensemble average.
% >>>
\subsection{Varying the size of the coarse-graining boxes}% <<<
\label{sec:size}

Before presenting the temporal and spatial dependence of the stress correlator,
let us investigate how the side length~$a$ of the averaging box changes the
fluctuations of the coarse-grained stress tensor inside.
\begin{figure}[bt]
  \centering
  \includegraphics{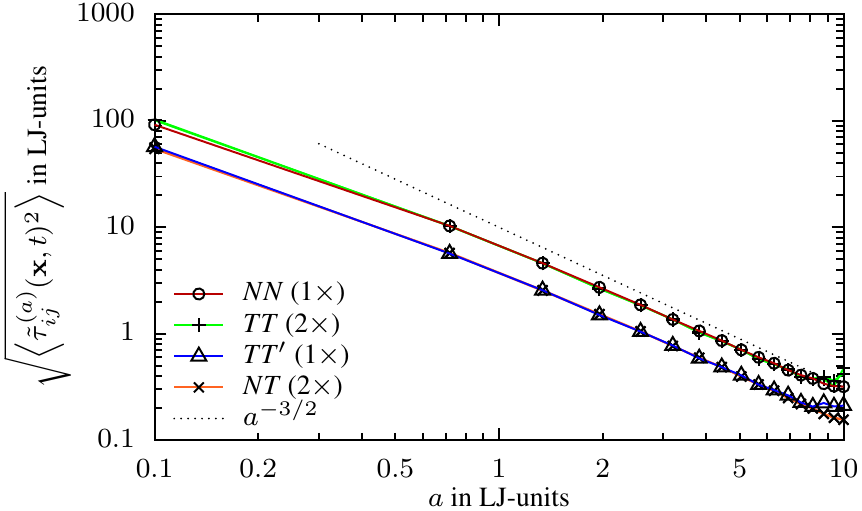}%
  \caption{(Colour on-line) Fluctuations of the stress tensor in bulk, as a
  function of the side~$a$ of the averaging box. All tensor components are
  plotted. They are grouped according to the expected symmetry (see text).}%
  \label{fig:size}%
\end{figure}
\begin{figure}[bt]
  \centering
  \includegraphics{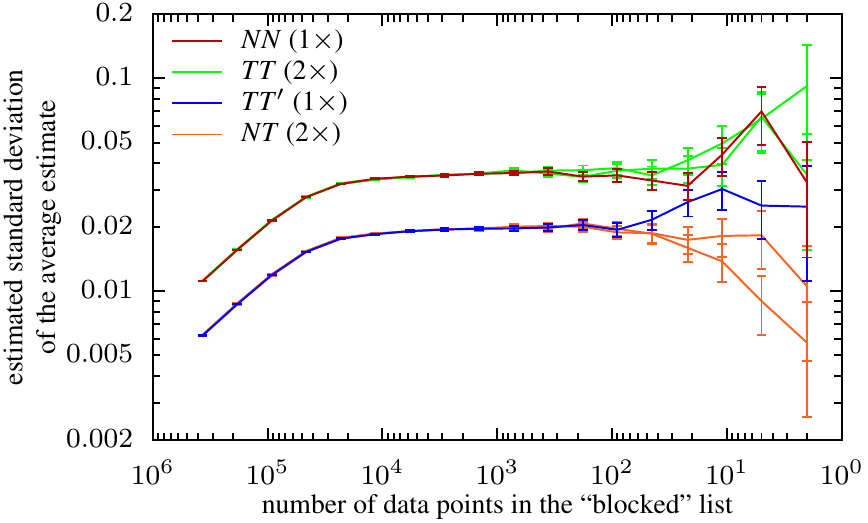}%
  \caption{(Colour on-line) Estimated error of the average estimation from the
  blocking transformation~\cite{FlyPet89}. We used the time series of all stress
  components in a box of size~$a=1.0$ centred between the plates.}%
  \label{fig:blocking}%
\end{figure}%

The explicit coarse-graining method by averaging over small cubes allows to
observe the transition from thermodynamic behaviour (in large boxes) to highly
fluctuating random behaviour (in small boxes). The amount of fluctuations
depends of course on the size of the box. As we treat here only
thermodynamically equilibrated fluids, the mean square deviation of the stress
tensor should scale as the inverse volume of the averaging box -- just as for
any intensive quantity~\citep[\textsection~112]{LanLif01}. In the bulk fluid, we
should thus find the deviations of the stress tensor to be
\begin{equation}
  \label{volscaling}
  \sqrt{\bigl\langle\RStress_{ij}(\bfx, t)^2\bigr\rangle} \propto a^{-3/2},
\end{equation}
where $\RStress_{ij}(\bfx,t)$ is defined by subtracting the macroscopic stress tensor,
which is $-p_0\delta_{ij}$,
\begin{equation}
  \label{RStress}
  \RStress_{ij}(\bfx,t) := \CGStress_{ij}(\bfx,t) + p_0\delta_{ij}.
\end{equation}
The value for the pressure is taken from disordered fluid atoms only, and not
from the layers close to the plates.

The numerical verification of the scaling~\eqref{volscaling} is plotted in
Fig.~\ref{fig:size}. Amazingly, we find the correct scaling already for box
sizes of~$a\sim2$, that is a side length of the order of the average distance
between the fluid atoms. The scaling~\eqref{volscaling} is originally based on a
thermodynamic argument which requires that many particles are involved.
\emph{Instead, we here see the correct scaling already for very few particles}.
It appears that there is more truth in the thermodynamic argument than just the
interaction of many particles.

For visualising the tensor of stress fluctuations, we have chosen the following
strategy, which will be applied also to the following plots: The symmetry
requirements (B)~and (C) in Sec.~\ref{sec:symmetry} lead to components which are
equal only in a statistical sense. All components which may exhibit such
fluctuations are plotted by individual lines. The legend and the colour coding,
however, differentiate only between the groups of components which should be
equal on average. The number of lines actually plotted is given in parentheses.

The fact that there is no difference between the normal and the parallel
directions in the data of Fig.~\ref{fig:size} confirms our assumption that the
fluid behaviour in the middle between the walls is the same as if they were
absent. There, the diagonal and the off-diagonal parts fluctuate differently.
This is a consequence of the components being different functions of the two
viscosities (see also Fig.~\ref{fig:viscosities}).

Exemplary for all plotted variables, we show for the centred boxes that the
system is well equilibrated. Fig.~\ref{fig:blocking} visualises the blocking
transformation~\citep{FlyPet89} of the diagonal entries in the coarse-grained
stress tensor. A~plateau is clearly visible, which shows that the analysed time
series indeed correspond to a stationary stochastic process.
% >>>
\subsection{Stress correlations in bulk fluid}% <<<
\label{sec:bulk}

\begin{figure}[!bt]%
  \centering
  \includegraphics{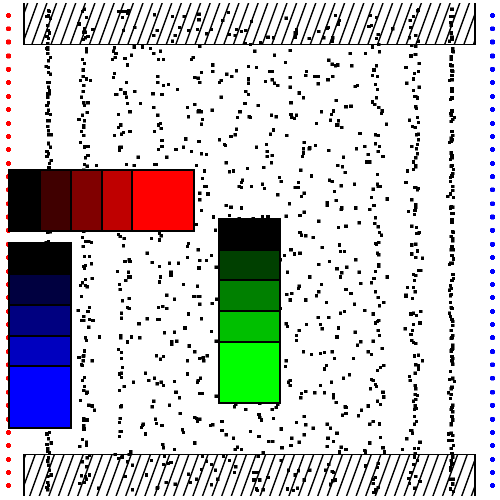}%
  \caption{(Colour on-line) Same as Fig.~\ref{fig:atoms}a, but with different
  series of overlapping coarse-graining cubes, which are used in
  Figs.~\ref{fig:platesbulk}--\ref{fig:platestemp}.}%
  \label{fig:boxes}%
\end{figure}

The aim of this paper is to determine the spatial and the temporal structure in
correlations of fluctuations of the coarse-grained stress tensor
$\RStress_{ik}$. We are interested in the difference between correlations in
bulk and in the near vicinity of the walls. In order to do so, we set out three
different series of cubic boxes, as depicted in Fig.~\ref{fig:boxes}. These
cubes are shifted by small distances and may overlap. The first series is in the
middle between the plates, where the fluid behaves as in bulk, a second one
close to a wall and parallel to it, and a third series is normal to the wall.
\begin{figure*}[bt]%
  \centering
  \includegraphics{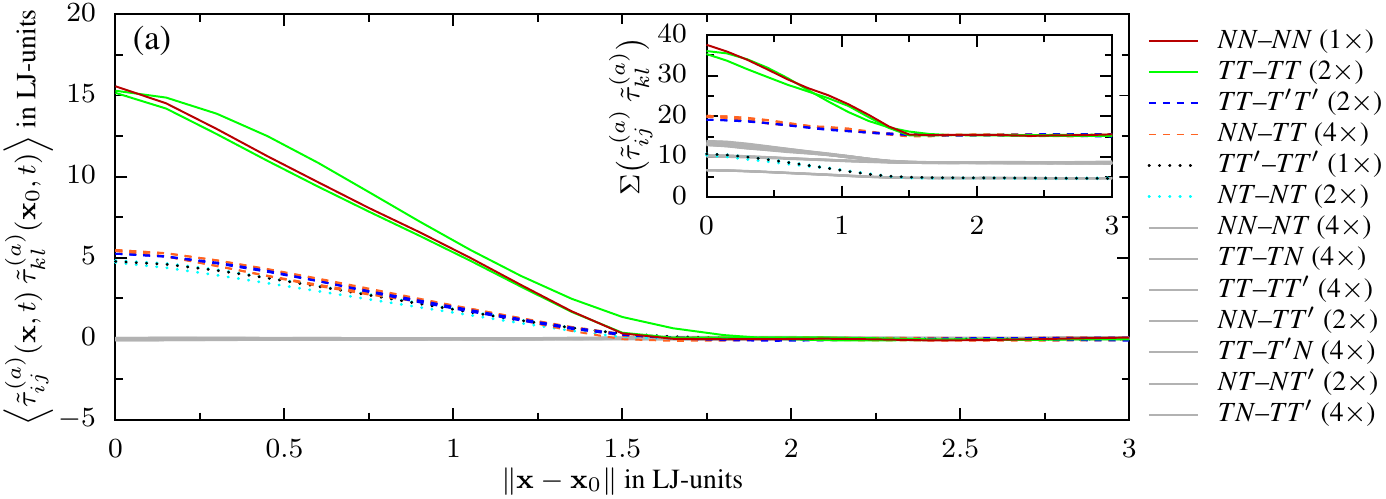}\\[1ex]
  \includegraphics{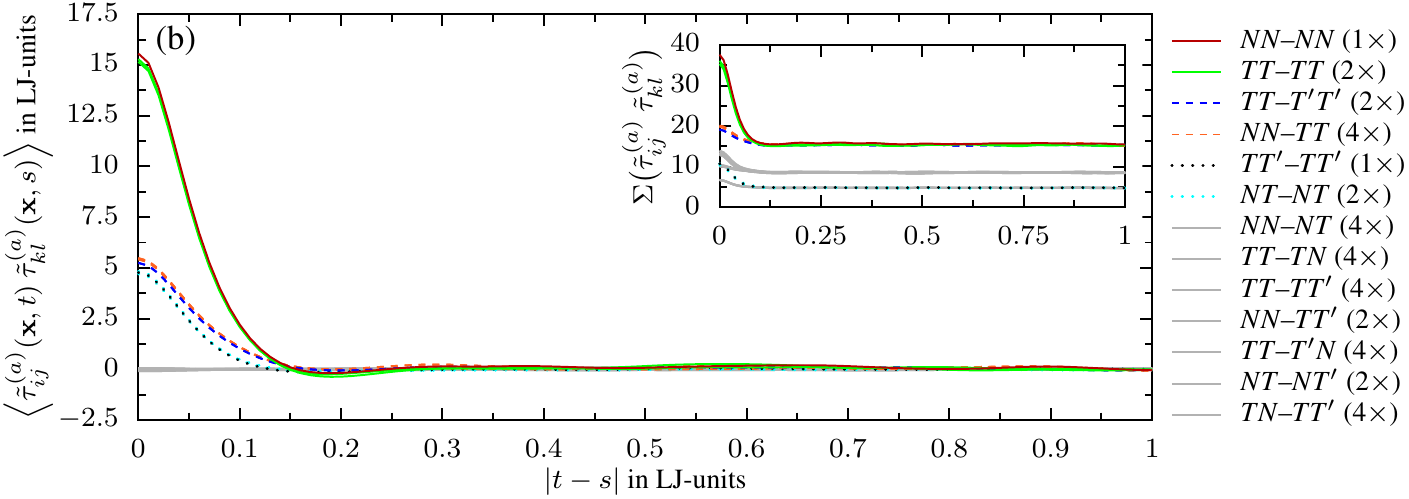}%
  \caption{(Colour on-line) Spatial and temporal correlations of the stress
  fluctuations in the middle of the domain (in bulk). Plotted are the average
  correlators and their standard deviations ($\Sigma(\cdot)$, see insets) for
  different classes of index combinations.}%
  \label{fig:platesbulk}%
\end{figure*}%

The displacement of the boxes makes it possible to determine the spatial form of
the stress correlation by cross-correlating the coarse-grained stress
fluctuations in different boxes. Temporal correlation functions will also be
given. We would like to compare the result with the uncorrelated theory in
Eq.~\eqref{bulkcorr}. However, the nonzero size of the boxes used for averaging
does not allow a direct comparison. Remind that $\RStress_{ij}$~depends on the
box size. We should find the delta function of Eq.~\eqref{bulkcorr} in the limit
$a\to0$, which is infeasible in practice. For all nonzero sizes, the numerical
data is sort of ``convoluted'' with the shape of the boxes. The best we can do
is to compare with the equally smoothed version of Eq.~\eqref{bulkcorr} using a
reasonably small~$a$. The corresponding correlator of the uncorrelated stress
tensor~$\ranstress_{ij}$ is expressed by the relative overlap of the two boxes,
\begin{equation}
  \label{gc_delta} \bigl\langle\RStress_{ij} \RStress_{kl} \bigr\rangle
  \propto \bigl|V^{a}(\bfx) \cup V^{a}(\bfy)\bigr|,
\end{equation}
of course disregarding the infinite term~$\delta(t{-}s)$ in
Eq.~\eqref{bulkcorr}. Here, $|\cdot|$~denotes taking the volume of a set.

The spatial and the temporal correlations of the random coarse-grained stress
tensor are depicted in Fig.~\ref{fig:platesbulk}. The data have been generated
by correlating the contents each box in the central series of
Fig.~\ref{fig:boxes} with the first one, which has $\bfx_0$ as its centre.
Plotted are all~$36$ components of the resulting rank-four tensor which can be
different using the symmetry~(A) of Sec.~\ref{sec:symmetry}. They are grouped
into the $13$~different classes according to the symmetries~(B) and~(C).
Fig.~\ref{fig:platesbulk} shows that seven of them vanish (grey lines) and that
the remaining six take three different values. This is expected according to
what has been said around Fig.~\ref{fig:viscosities}.

The spatial correlations in Fig.~\ref{fig:platesbulk}a should be compared with
the results expected from an uncorrelated stress tensor. These would be lines
starting at some value which is related to the viscosities (unspecified due to
$\delta(t{-}s)$ in Eq.~\eqref{bulkcorr}) and which goes to zero the distance~$a$
and beyond. The coincidence in Fig.~\ref{fig:platesbulk}a is remarkable. The
numeric values are a bit more smoothed, but the apparent convolution kernel
responsible for this smoothing has a width of less than unity. \emph{We
therefore conclude that Eqs.~\eqref{bulkcorr} with Eq.~\eqref{isotropy} in bulk
are very precise, down to a length scale of the order of the molecule distance.}
Similarly to what we found in Sec.~\ref{sec:size}, such a strong result cannot
be expected from thermodynamic considerations.

The temporal correlations in Fig.~\ref{fig:platesbulk}b exhibit the deviations
from the temporal delta function in Eq.~\eqref{bulkcorr}. The decay is
sufficiently fast to reasonably adopt the approximation of the delta function.
Whether there is an algebraic contribution, the so-called \emph{long-time tail}
cannot be judged from these data. In order to calculate the viscosity values, we
use the integral over the detailed time correlator, as proposed by the
Green--Kubo equation. The idea is the same as in
Refs.~\citep{RowPai97,MeiLaeKab04,MeiLaeKab05}, only that we take a box smaller
than the simulation box. For the shear viscosity, this reads
\begin{equation}
  \label{GreenKubo}
  \eta = \frac{a^3}{kT} \int\limits_0^\infty dt\:
  \Bigl\langle\RStress_\textit{NT}(\bfx,0)\:\RStress_\textit{NT}(\bfx,t)\Bigr\rangle,
\end{equation}
and similarly for the two components comprising the volume viscosity~$\lambda$.
From the data in Fig.~\ref{fig:platesbulk}b we find the values $\eta=0.84$ and
$\lambda=1.72$ by regression.\footnote{Notice that these values do not quite
correspond to the finding in Refs.~\cite{MeiLaeKab04,MeiLaeKab05}. A possible
explanation is the dependence of the viscosities on the size~$a$ of the
averaging box. We found such a dependence together with long-time tails in the
autocorrelation of the random stress tensor in bulk.}
% >>>
\subsection{Stress correlations near a rigid flat wall}% <<<
\label{sec:plate}

Near the plates, the stress correlations look quantitatively and qualitatively
different. The average stress tensor and also its fluctuations are anisotropic.
Our aim in this section is to determine the five~\emph{surface viscosities} in
order to quantify their deviations from the bulk values, according to the scheme
in Fig.~\ref{fig:viscosities}.%
\begin{figure}[bt]
  \centering
  \includegraphics{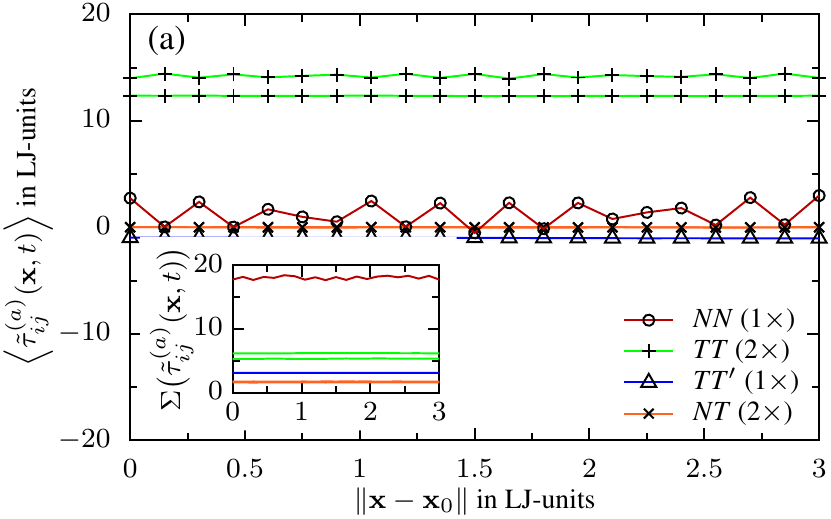}\\
  \includegraphics{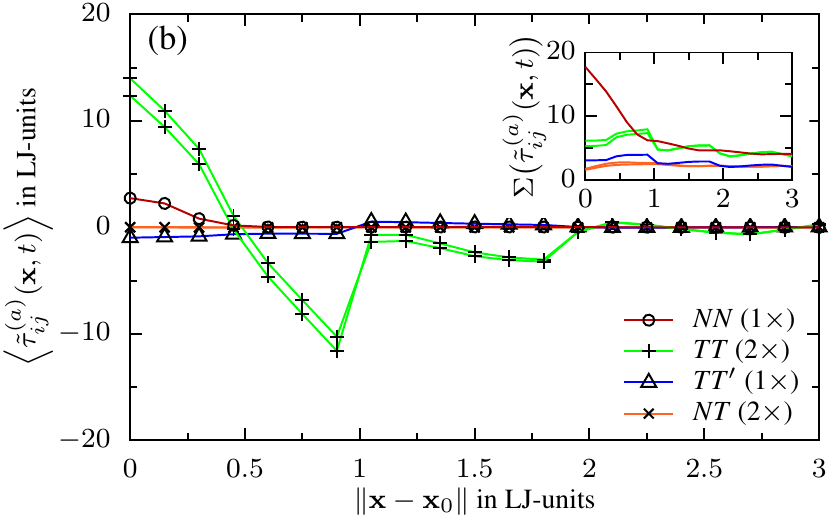}%
  \caption{(Colour on-line) The coarse-grained stress tensor as a function of
  the position, as indicated in two series close to the wall in
  Fig.~\ref{fig:boxes}: (a)~parallel to the plate, (b)~orthogonal to the
  plate.}%
  \label{fig:platesstress}%
\end{figure}%

First, let us look at the coarse-grained random stress tensor itself.
Fig.~\ref{fig:platesstress} reveals the mechanical properties of the first fluid
layers: We find a huge lateral diagonal stress (\textit{TT}~components) which
strongly depends on the distance to the wall but not on the lateral position. In
the figure, the points~$\bfx$ correspond to the centres of the series of boxes
laid out in Fig.~\ref{fig:boxes}. The position~$\bfx_0$ is the centre of the
first box in the normal series, which is closest to the wall. The average value
of the huge diagonal stress is even larger than its fluctuations. This lateral
stress can be understood as a \emph{surface tension} of the contact between
liquid and solid. The \textit{NN}-component is much smaller---its average value
is the same as the bulk pressure. The \textit{NN}-component is dominated by its
fluctuations which grow in the vicinity of the wall and even overgrow the
tangential stresses. In the following, we will find that there are correlations
in these fluctuations. The lines in Fig.~\ref{fig:platesstress}b prove that both
the average values and the standard deviations are independent of the lateral
position. The roughness of the \textit{NN}-component in this plot is due to the
large fluctuations.

For the data in Fig.~\ref{fig:platesstress} and in the following ones, another
comment on the symmetry is necessary. By the noted symmetries in
Sec.~\ref{sec:symmetry}, the two lines for the \textit{TT}-components in
Fig.~\ref{fig:platesstress} should coincide. However, they do not. The reason is
that the structured liquid is not entirely isotropic in the tangential
directions, but it is ordered in a hexagonal grid, as depicted in
Fig.~\ref{fig:atoms}b. This strict ordering is a result of the wall being very
flat. A~roughness of the size of one atom distance would easily destroy the
hexagonal structure, and only the symmetries of Sec.~\ref{sec:symmetry} would
remain. In order to weaken the effect, we averaged over eight~trajectories which
differed only by the orientation of the hexagonal grid. This reduced the
deviations considerably to those in Fig.~\ref{fig:platesstress}.
\begin{figure*}[bt]
  \centering
  \includegraphics{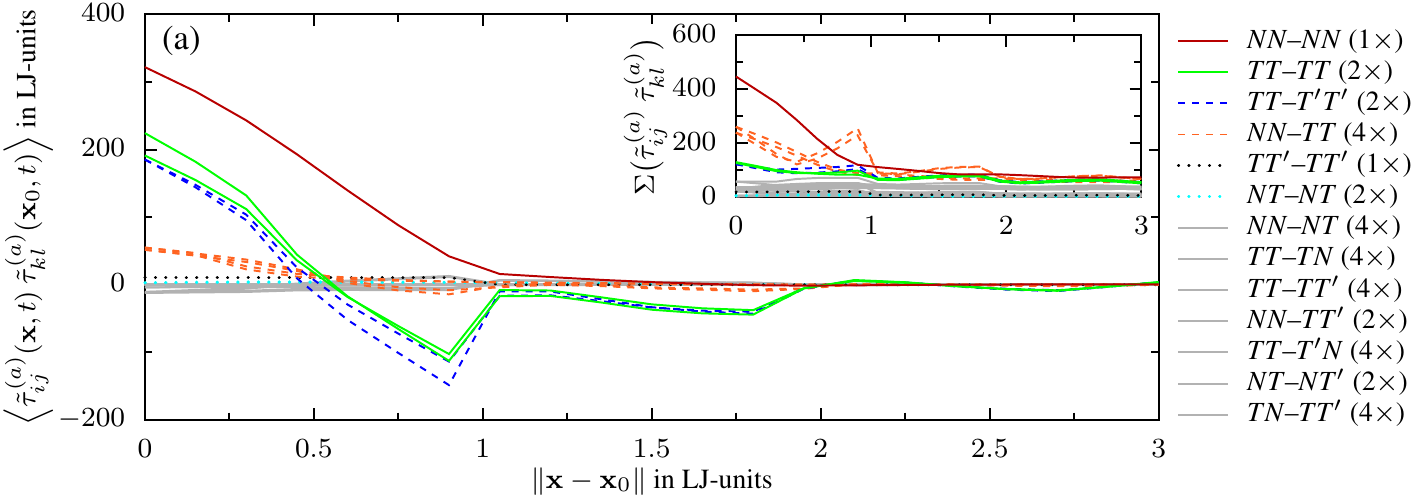}\\[1ex]%
  \includegraphics{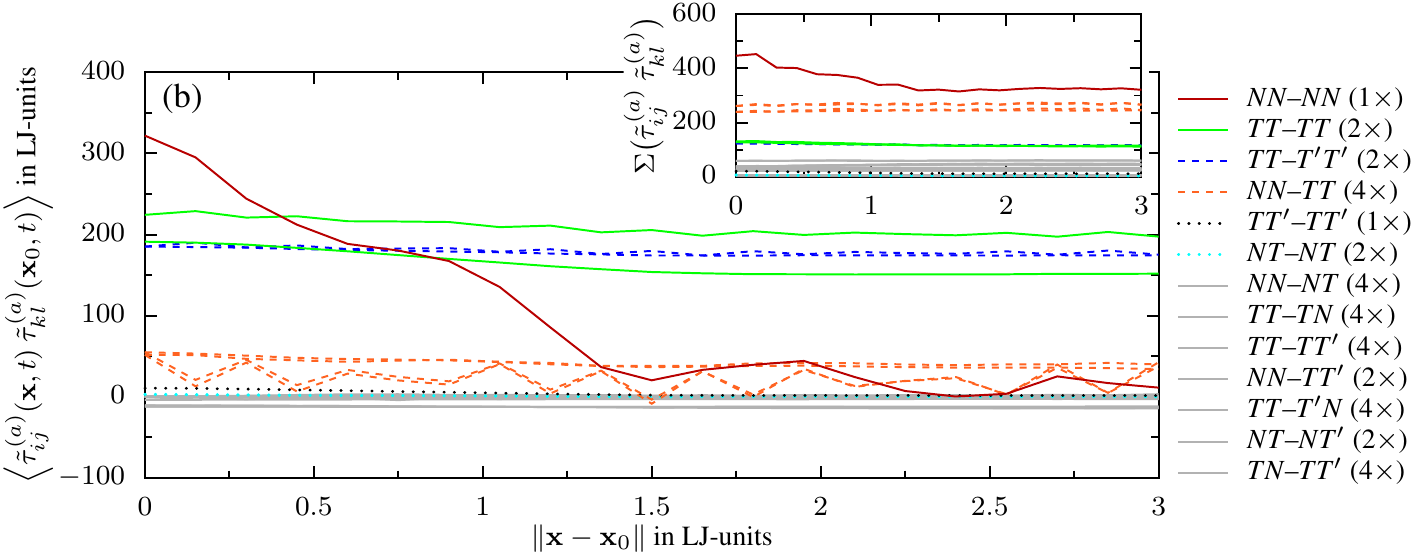}%
  \caption{(Colour on-line) Spatial correlations of the random stress tensor in
  the vicinity of a wall. (a) in normal direction, and (b) parallel to the wall,
  see the series of cubes in Fig.~\ref{fig:boxes}. Each box in the series is
  correlated with the first one, centred around~$\bfx_0$ (in the orthogonal
  case: the one closest to the wall). Legend and inset as in
  Fig.~\ref{fig:platesbulk}a.}%
  \label{fig:platesdist}%
\end{figure*}
\begin{figure*}[bt]
  \centering
  \includegraphics{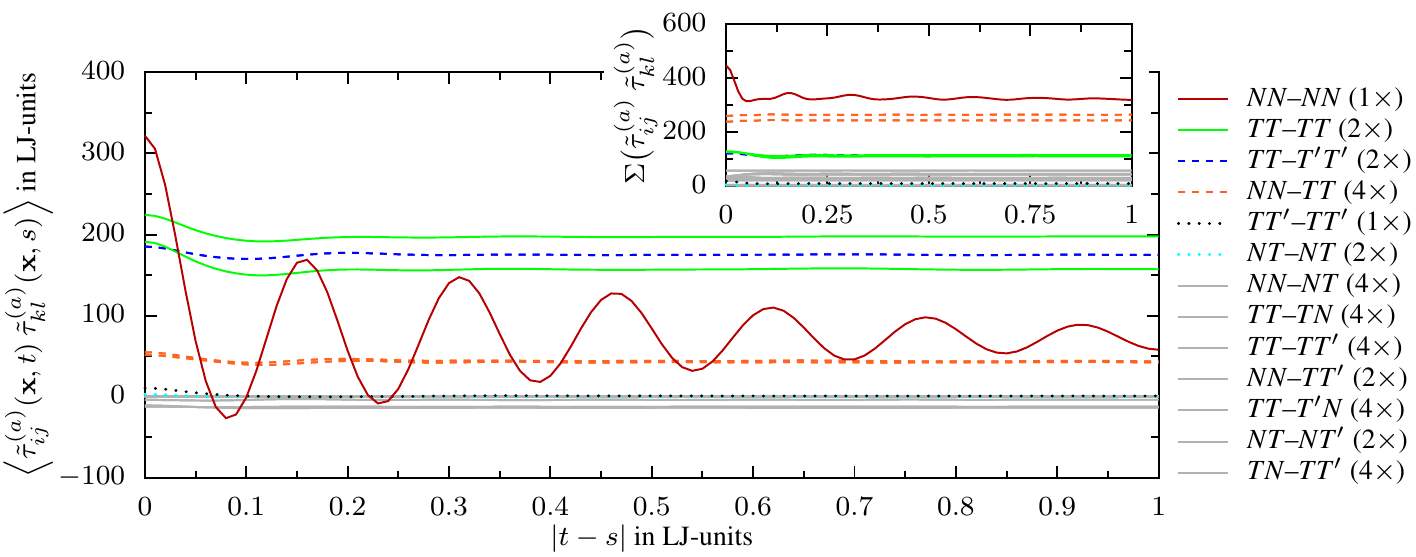}%
  \caption{(Colour on-line) Temporal correlations of the random stress tensor
  near a wall. Legend and inset as in Fig.~\ref{fig:platesbulk}b.}%
  \label{fig:platestemp}%
\end{figure*}

The spatial correlations of the random stress tensor are depicted in
Fig.~\ref{fig:platesdist}. Again, the two series of integration boxes in
different directions are used, see Fig.~\ref{fig:boxes}. The plot shows the
correlation of each cube in the series with the first one, which is in the
normal series the one closely aligned at the wall. For the normal series, the
result is not expected to be independent of this choice. Only in the parallel
series, the correlation should depend only on the distance of the boxes. As
explained above, there are in principle 13~different results to be expected.
Seven of them, those indicated in grey, are zero. The six non-zero components
are plotted in colour. The correlations of the stress tensor inherit the main
features of the stress in Fig.~\ref{fig:platesstress}, in particular the
systematic part which we called a surface tension. In normal direction, the
tangential components exhibit a similar anti-correlation as in
Fig.~\ref{fig:platesstress}b. The decay of the most pronounced correlations
takes place on the length of one atom-distance and is independent of the box
size~$a$. We verified this statement using other box sizes. In tangential
direction, the surface tension manifests itself in a nonzero value of the
tangential correlation, see Fig.~\ref{fig:platesstress}b.

Apart from the surface tension, we find in Fig.~\ref{fig:platesstress}b a
similar behaviour as in bulk, that is a linear decay up to a distance~$a$ and
zero beyond. This is clearly visible for the \textit{NN}--\textit{NN} component,
but also for the others---after subtracting the contribution of the surface
tension. This implies that the spatial delta function still works parallel to
the walls. The correlation in normal direction, which is dominated by the
details of the interaction between wall and fluid, depends both on the distance
between the box and their distance to the wall. These two considerations lead us
to the following proposition for the correlation function of the random stress
tensor near an immobile flat rigid wall:
\begin{multline}
  \label{surfcorr}
  \bigl\langle\ranstress_{ij}(\bfx,t)\,\ranstress_{kl}(\bfy,s)\bigr\rangle
  = 2kT\,A_{ijkl}\,\delta(t{-}s) \times{}\\
    \delta^2(\bfx^{||}{-}\bfy^{||})\,f(\bfx^\bot,\bfy^\bot).
\end{multline}
with the tensor~$A_{ijkl}$ from Eq.~\eqref{surfiso}, depending on the surface
viscosities, and with $\bfx^{||}$ and $\bfx^{\bot}$ the parallel and normal
projections, respectively. The function~$f$, which has the physical dimension of
inverse length, takes into account the microscopic details of the interaction
between wall and fluid. As the only length scale involved is the distance
between wall and the first structured liquid layer, we expected $f$~to scale as
one over this distance.

It remains to check whether the delta function in time is still a good
approximation, and what values the surface viscosities have. The temporal
correlations near the wall are depicted in Fig.~\ref{fig:platestemp}. The
results are taken from one of the boxes in the parallel series. Here, we see
even better than in Fig.~\ref{fig:platesdist} that there is a systematic
contribution coming from the surface tension. Notice that it affects only the
tangential components and not the heavily oscillating \textit{NN}--\textit{NN}
component, which is decaying to zero, but on a longer time scale than plotted.

In order to obtain the viscosities from these data, we have subtracted the
asymptotic value from the correlation function, which renders the function
integrable. This means that we implicitly subtracted a sort of local pressure
which is allowed to be anisotropic. This pressure reflects the static
contributions from the interaction liquid--wall, which we called surface tension
above. The Green--Kubo integral of the such normalized correlation functions
yields the following values for the surface viscosities:
\begin{equation}
  \label{surfviscosities}
  \begin{aligned}
   &\eta_1 \approx 113, \\
   &\text{$\eta_2$ between 16 and 34}, \\
   &\eta_3 \approx 8.4, \\
   &\text{$\eta_4$ between 65 and 178}, \\
   &\text{$\eta_5$ between 4.2 and 5.5}.
  \end{aligned}
\end{equation}
The values for these viscosities are not very precise for three reasons: The
first is the hexagonal structure of the first fluid layer which does not
coincide with the Cartesian coordinates used for the tensor indices. We thus
expect different viscosity values for different orientations of the hexagonal
grid. The above mentioned average over eight different trajectories with
different orientations apparently does not suffice to obtain unified values. The
second reason is that the thermodynamic pressure is not isotropic anymore and
that the time-correlation function does not decay to zero. We thus had to strip
off a linear time-dependency before integrating, as already explained. The third
reason is that the intrinsic time scales of the time-correlation function are
much longer than in bulk. While the bulk viscosities could be easily identified
from a Green--Kubo integral over a few dozens of timesteps, it required thousand
and more timesteps to do the same for the surface viscosities. This much longer
time scale can already be guessed from the comparison of
Figs.~\ref{fig:platesbulk}b and~\ref{fig:platestemp}.
% >>>
\subsection{Hydrophobic walls}% <<<

The rigid plates used in the above simulations consist of the same type of atoms
as the liquid. Such walls are known to be hydrophilic, which is also seen in the
strict ordering of the first structured fluid layer. We repeated the simulation
using hydrophobic walls, which have a modified Lennard--Jones potential to
interact with the fluid atoms~\cite{BarBoc99},
\begin{equation}
  \label{LJphobic}
  \phi(r) = 4\epsilon\bigl[(\sigma/r)^{12} - c (\sigma/r)^6\bigr]
\end{equation}
with the same cutoff and with an equivalent smoothing of the potential as in
Eq.~\eqref{LJsmooth}. According to \citet{BarBoc99}, who proposed the
form~\eqref{LJphobic} for simulating hydrophobic walls, the factor
factor~$c=0.5$ which we used, should correspond to a contact angle of about
$140^\circ$.

The simulation leads to a similarly structured liquid as in
Fig.~\eqref{fig:atoms}, only that the first layers are a bit farther away from
the walls, that they are not as flat, and that their lateral structure is not as
regular. The ordering effect of the first layers are nevertheless very strong,
and must therefore be induced by the flat geometry of the wall. We took the same
bulk pressure and density as in the previous simulation. As expected for a
hydrophobic surface, the \textit{TT}-components in Fig.~\ref{fig:platesstress}
are different. They are considerably smaller, around the value~$3$ instead
of~$13$. The correlation functions look similar to those depicted in
Figs.~\ref{fig:platesdist} and \ref{fig:platestemp}, with the difference that
they decay much faster in time, in particular the \textit{NN}--\textit{NN}
component. Accordingly, also the surface viscosities turn out to be different.
Near the hydrophobic surfaces, we find the values
\begin{equation}
  \label{surfviscosities_phobic}
  \begin{aligned}
   &\eta_1 \approx 87, \\
   &\eta_2 \approx 1.7, \\
   &\eta_3 \approx 1.5, \\
   &\eta_4 \approx 8.1, \\
   &\eta_5 \approx 0.35.
  \end{aligned}
\end{equation}
It is not astonishing that these viscosities are smaller than those found near
the wetting walls. In particular in tangential direction, where the liquid is
feels less constraints, the dissipation is smaller. This affects
$\eta_2$--$\eta_5$, but not $\eta_1$, which is accordingly of a similar
magnitude as in the wetting-wall case. Particularly astonishing is the small
value of~$\eta_5$, which takes over the role of the shear viscosity. It is much
smaller than the bulk value ($\eta = 0.84$ also in the hydrophobic simulation).
% >>>

\section{Discussion}% <<<
\label{sec:discussion}

The numerical observations above, as described in Sections~\ref{sec:bulk} and
\ref{sec:plate} allow to estimate the validity of the main assumptions leading
to fluctuating hydrodynamics. Indeed, we observed that the spatial delta
function in Eq.~\eqref{bulkcorr} works very well in bulk. Near a plate, which
has been assumed to be flat, immobile and rigid, the isotropy is restricted to
rotations parallel to the wall, and we must replace the spatial delta function
by one which is only acting parallel to the wall, as given in
Eq.~\eqref{surfcorr}. In orthogonal direction, the details of the dynamical
interaction between liquid and wall introduce an unknown factor in form of a
function depending on the orthogonal distances. Moreover, we have to deal with
five viscosities instead of two, the values of which are provided in
Eq.~\eqref{surfviscosities}.

The parallel structure of the spatial delta functions and the number of
viscosities are both imposed by pure symmetry and cannot be disputed. But we may
ask which of the surface viscosities will have an effect on measurable
quantities, such as on the flow profile in nanopores or on the motion of a
Brownian particle. If we regard only forces exerted on the rigid walls---and
their correlation functions---, there are only two viscosities which may play a
role, namely those possessing a normal direction in both contributing stresses.
They are easily identified in the scheme in Fig.~\ref{fig:viscosities} to be
$\eta_1$~and $\eta_5$, the latter of which takes over the role of the bulk shear
viscosity, and the first one the role of the bulk volume viscosity. It is
interesting to see that the value of~$\eta_5$ is entirely independent on the
shear viscosity in bulk. It depends more on the hydrophobic/hydrophilic
character of the walls. The surface-equivalent of the volume viscosity,
$\eta_1$, is in both cases nearly two orders of magnitude larger than in bulk.

The finding that the surface viscosities do not have values comparable to the
bulk viscosities brings us now to more profound questions on the dynamical
theory involved: For the Brownian particle, we argue that only the first
surrounding liquid layers do actually interact with the particle. There, the
relevant dissipation coefficients are not the bulk viscosities but the two
viscosities $\eta_1$~and $\eta_5$. It appears thus as a reasonable to ask why we
do not see~$\eta_5$ instead of~$\eta$ appearing in the Stokes--Einstein formula
for the diffusion coefficient? We have to leave this question open at the
moment. A possible answer might go in the same direction as the discussion of
time correlations and spatial ordering effects of Ref.~\citep{MokYulHan05}.

A possible answer is certainly linked to the fact that the surface viscosities
here are \emph{local} quantities. They have been determined using
coarse-graining over very small boxes. The usual definition of a ``viscosity'',
however, implies limits to large boxes and long time integrals (or the limit
$\omega\to0$ after the limit $\bfk\to\bfzero$ in Fourier space and
time~\citep[p.~186]{KubTodHas85}). In this limit, the five surface viscosities
cannot be identified anymore. Another implication of the use of \emph{local
viscosities}, which are not viscosities in the usual sense of a large-box limit,
is that they do not lead to differential equations. They rather appear as
integral kernels, nonlocal both in space and in time, in integro-differential
equation. We cannot say much about the possibility to find an to interpret such
equations at the moment, the less as we expect the long-time tails in the
autocorrelation function, thus nonlinear terms, to intervene. Nevertheless, we
would like to stress that we do not expect the qualitative nature of the
anisotropy described here---that is the large difference of surface and bulk
viscosity values, and that the surface viscosities do not depend on the bulk
viscosities---to depend on the precise temporal behaviour. The long-time tails
appearing in the correlation functions should only change the quantitative
values of the viscosities, leaving the presented material untouched.
% >>>

\section*{Acknowledgements} % <<<

I would like to dedicate this paper to Peter Hänggi, on occasion of his 60th
birthday, which is celebrated by the present special issue of Chemical Physics.
The subject of this paper, in particular the aspects explained in the last
section touches in part his early works on stochastic processes with memory.

I would further like to express my thanks to Peter Talkner for waking my
interest in fluctuating hydrodynamics, and to Laurent Joly and Lydéric Bocquet
for introducing me to the molecular dynamics package~\LAMMPS. Further sincere
thanks go to Anthony C.~Maggs for pointing me to the blocking method and to
Claire Lemarchand, Ken Sekimoto, Falko Ziebert and Jörg Rottler for helpful
discussions.
% >>>

%\bibliographystyle{elsarticle-num-names}
%\bibliography{paper}
% <<<

% >>>
\end{document}